\documentclass{article}

\usepackage{microtype}
\usepackage{graphicx}
\usepackage{subfigure}
\usepackage{booktabs} 
\usepackage{algorithm}
\usepackage{algorithmic}
\usepackage{adjustbox}
\usepackage{float} 
\usepackage{fancyvrb} 

\usepackage{hyperref}

\usepackage[accepted]{icml2025}

\usepackage{amsmath}
\usepackage{amssymb}
\usepackage{mathtools}
\usepackage{amsthm}

\usepackage[capitalize,noabbrev]{cleveref}

\theoremstyle{plain}

\theoremstyle{definition}

\theoremstyle{remark}

\usepackage[textsize=tiny]{todonotes}

\icmltitlerunning{DeepSeq — Agentic Generative AI Foundation Models for Structured Single-Cell RNA Sequencing Data}

\begin{document}

\twocolumn[
\icmltitle{DeepSeq: High-Throughput Single-Cell RNA Sequencing Data Labeling \\ via Web Search-Augmented Agentic Generative AI Foundation Models}



\icmlsetsymbol{equal}{*}

\begin{icmlauthorlist}
\icmlauthor{Saleem A. Al Dajani}{mit}
\icmlauthor{Abel Sanchez}{mit}
\icmlauthor{John R. Williams}{mit}

\end{icmlauthorlist}

\icmlaffiliation{mit}{Department of Civil and Environmental Engineering, Massachusetts Institute of Technology, 77 Massachusetts Avenue, Cambridge, MA, 02139, United States of America}

\icmlcorrespondingauthor{Saleem A. Al Dajani}{sdajani@mit.edu}

\icmlkeywords{Agentic AI, large language models (LLMs), single-cell RNA sequencing}

\vskip 0.3in
]



\printAffiliationsAndNotice{}  

\begin{abstract}

Generative AI foundation models offer transformative potential for processing structured biological data, particularly in single-cell RNA sequencing, where datasets are rapidly scaling toward billions of cells. We propose the use of agentic generative AI foundation models with real-time web search to automate the labeling of experimental data, achieving up to 82.5\% accuracy. This addresses a key bottleneck in supervised learning for structured omics data by increasing annotation throughput without manual curation and human error. Our approach enables the development of virtual cell foundation models capable of downstream tasks such as cell-typing and perturbation prediction. As data volume grows, these models may surpass human performance in labeling, paving the way for reliable inference in large-scale perturbation screens. This application demonstrates domain-specific innovation in health monitoring and diagnostics, aligned with efforts like the Human Cell Atlas and Human Tumor Atlas Network.

\end{abstract}

\section{Introduction \& Background}

Single-cell RNA sequencing (scRNA-seq) has transformed our ability to understand biological systems at cellular resolution, enabling the decomposition of heterogeneous tissues into interpretable cellular subpopulations \cite{hicks2018bioinfosummer,shalek2013single}. Unlike bulk sequencing, which averages gene expression across thousands of cells, single-cell approaches preserve cellular diversity and support downstream analyses such as lineage tracing, perturbation inference, and cell-type identification.

A key challenge emerging from these advances is scale. With improved protocols and barcoding methods, scRNA-seq datasets have grown from thousands to millions of cells per experiment, opening the door to system-level modeling of gene regulation and cellular behavior. However, the complexity and dimensionality of these datasets far outpace manual annotation methods, particularly as the number of clusters grows with data volume. This challenge becomes even more pressing when considering tasks like supervised learning, pseudotime ordering, and perturbation modeling, which rely on accurate and interpretable cell-type labels.

\begin{figure}[!h]
    \centering
    \includegraphics[width=0.9\linewidth]{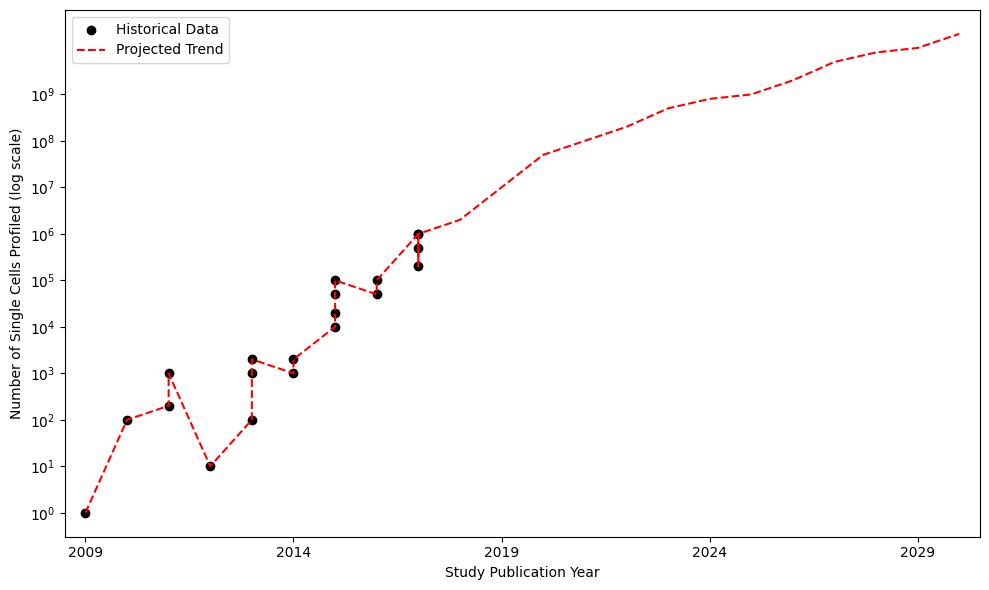}
    \caption{Exponential growth of single-cell sequencing enables foundation model–scale datasets. The number of single cells profiled per study has followed an exponential trend since 2009, resembling Moore’s law and enabling projections exceeding $10^9$ cells by 2030. Historical data points (black dots) are shown alongside a log-scale projection (red dashed line), adapted from \cite{svensson2018exponential}. This scaling trend motivates the development of foundation models tailored to structured single-cell data, which require billion-scale inputs for training on tasks such as annotation, perturbation modeling, and virtual cell simulation.}
    \label{fig:scaling-projection}
\end{figure}

In this work, we introduce \textbf{DeepSeq}, \url{https://github.com/saleemaldajani/deepseq} a pipeline that applies large language models (LLMs) to automate labeling of structured single-cell data using top marker genes from unsupervised clustering. DeepSeq supports both local inference using lightweight models and agentic web-enhanced querying via GPT-4o. The system is designed for reproducibility and scalability, incorporating filtering, dimensionality reduction, structured prompt generation, and accuracy benchmarking. 

In the following sections, we describe the DeepSeq architecture and algorithms in detail, demonstrate its annotation accuracy across multiple LLM configurations, and discuss implications for high-throughput cell atlas construction and virtual cell modeling.

\begin{figure}[!h]
    \centering
    \includegraphics[width=\linewidth]{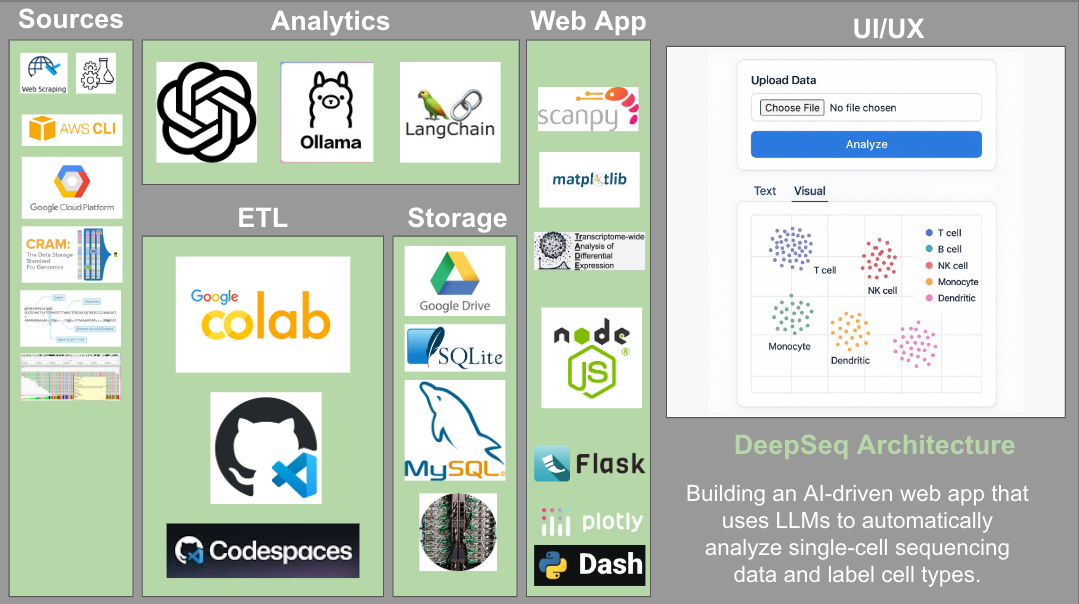}
\caption{DeepSeq system architecture. DeepSeq is a foundation model–powered web application designed for automated labeling of structured single-cell RNA sequencing data. The pipeline integrates large language models (OpenAI, Ollama), orchestration frameworks (LangChain), and differential expression analysis tools (Scanpy, TRADE) to process high-throughput omics data. It spans ETL, analytics, and deployment layers using cloud-native platforms (Colab, Codespaces) and standardized biological formats (CRAM, H5AD). DeepSeq demonstrates a domain-specific application of foundation models for scalable biomedical data annotation and virtual cell modeling.}
    \label{fig:architecture}
\end{figure}

\section{Methods}

The DeepSeq pipeline integrates single-cell RNA-seq preprocessing with foundation model–driven cell-type annotation using large language models (LLMs). The full workflow spans filtering, clustering, marker gene extraction, prompting, and structured evaluation. All core analysis and evaluation scripts are provided in the public repository.

\begin{algorithm}[H]
\caption{LLM-Based Cell-Type Labeling with DeepSeq}
    \label{alg:labeling}
\begin{algorithmic}[1]
\REQUIRE Filtered gene expression matrix $X$ with clusters $C_1, C_2, \dots, C_k$
\ENSURE Predicted cell-type label $\hat{y}_i$ for each cluster $C_i$
\FOR{each cluster $C_i$}
    \STATE Identify top marker genes $G_i = \texttt{rank\_genes}(C_i)$
    \STATE Construct prompt $P_i \gets$ format($G_i$)
    \IF{using \texttt{Ollama}}
        \STATE $\hat{y}_i \gets \texttt{local\_LLM}(P_i)$
    \ELSIF{using \texttt{GPT-4o}}
        \STATE Perform web search via OpenAI Agent
        \STATE $\hat{y}_i \gets \texttt{gpt4o}(P_i, \text{web results})$
    \ENDIF
    \STATE Evaluate $\hat{y}_i$ via marker match and label accuracy
\ENDFOR
\end{algorithmic}
\end{algorithm}

\subsection{Preprocessing and Filtering}

Raw single-cell data is processed into gene-by-cell matrices and converted into the \texttt{AnnData} format. Filtering is performed using three strategies: (1) standard thresholding (e.g., $\geq$200 genes per cell), (2) automated knee-point detection using \texttt{KneeLocator}, and (3) smoothed inflection-based filtering. These methods produce cleaned datasets with visual diagnostics for quality control.

\subsection{Clustering and Marker Gene Extraction}

Dimensionality reduction is performed using PCA, and cells are clustered using the Leiden algorithm based on neighborhood graphs. UMAP is used to embed cells in 2D for visualization. For each cluster, the top marker genes are identified using Scanpy’s ranking functions and are used to construct structured prompts for LLMs.

\subsection{LLM-Based Annotation}

LLMs are prompted with top-ranked marker genes per cluster to generate candidate cell-type labels. DeepSeq supports both local inference (via \texttt{Ollama}) and agentic inference (via \texttt{gpt-4o} with web search). Prompt orchestration and postprocessing are handled by LangChain. Prompts are designed following the format described by Hou and Ji \cite{hou2024assessing}, adapted to structured transcriptomic data.

\begin{figure}[!h]
    \centering
    \includegraphics[width=\linewidth]{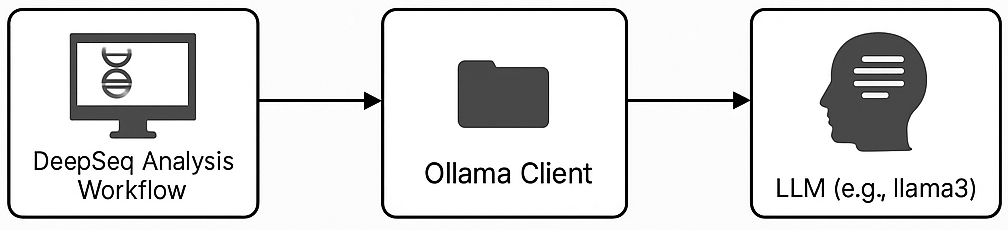}
        \includegraphics[width=\linewidth]{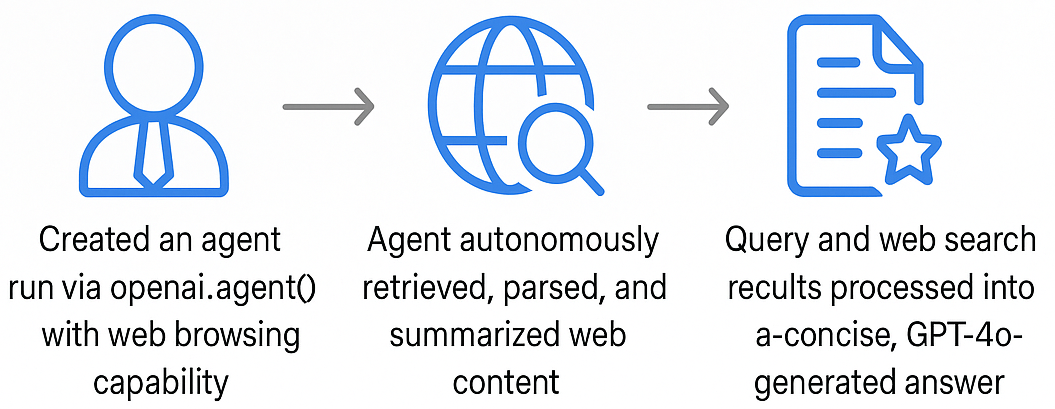}
    \caption{Dual inference workflows in DeepSeq for structured data annotation. The top panel illustrates local inference via the Ollama client, enabling efficient deployment of domain-specialized LLMs (e.g., LLaMA3) for on-device cell type labeling. The bottom panel depicts a live agentic inference pipeline using GPT-4o with web search capabilities, where an OpenAI agent autonomously retrieves and summarizes external content to augment biological annotations. Together, these workflows demonstrate the versatility of foundation models in structured biomedical pipelines under both offline and online settings.}
    \label{fig:llm-workflows}
\end{figure}

\subsection{Label Evaluation and Ground Truth Assessment}

To evaluate the precision of LLM-based labeling, we implement a two-stage validation protocol: \vspace{-0.5em}

\begin{itemize}
    \item \textbf{Marker Gene Verification:} We confirm that the top marker genes per cluster sufficiently match known canonical markers for each predicted label, ensuring that the evaluation is biologically meaningful. \vspace{-0.5em} 
    
    \item \textbf{Label Accuracy Assessment:} We compute the accuracy of LLM-generated labels by comparing them to manually curated ground truth labels. The comparison accounts for fuzzy string matching and synonym resolution to robustly assess agreement at the cluster level. \vspace{-0.5em} 
\end{itemize}

This framework ensures reproducible, interpretable evaluation of foundation models in structured single-cell data domains.

\begin{figure}[!h]
    \centering
    \includegraphics[width=\linewidth]{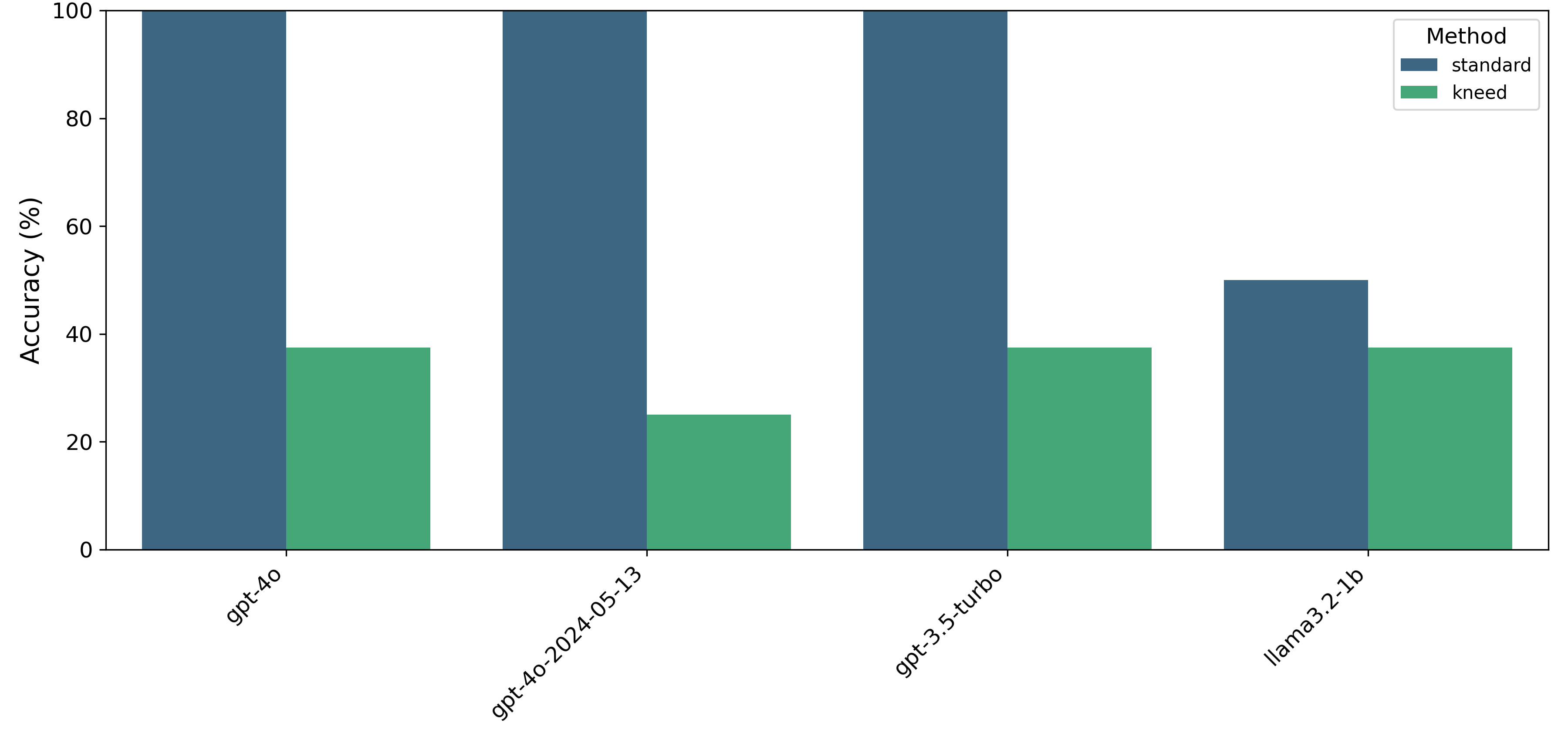}
    \includegraphics[width=\linewidth]{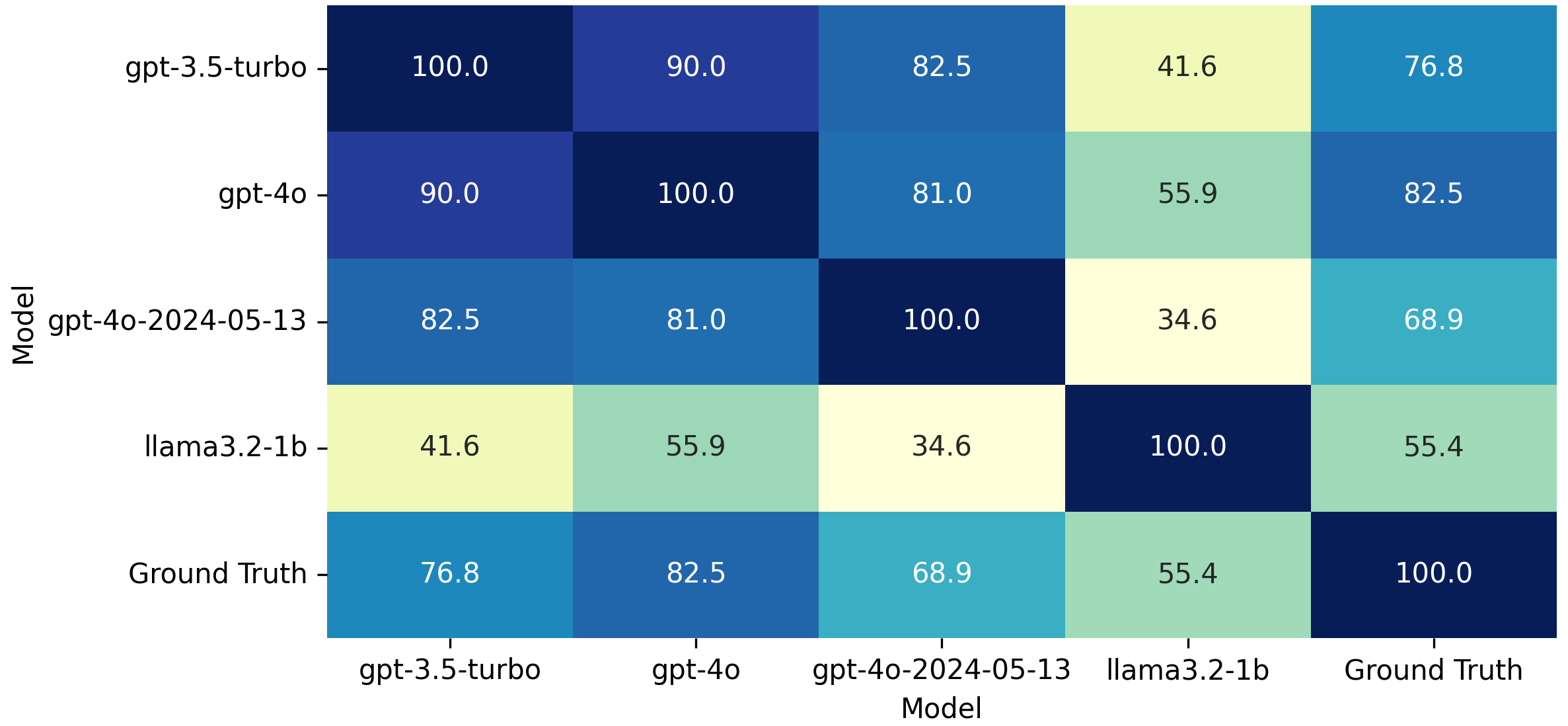}
\caption{Evaluation of foundation model accuracy in cell-type labeling from structured single-cell data. Top panel: Agreement between top marker genes per cluster and a ground truth reference ensures the validity of downstream label evaluations. Bottom panel: Accuracy comparison across LLMs using cluster-level marker gene inputs. The evaluation method was automated through prompting strategies inspired by \cite{hou2024assessing}. The agentic \texttt{gpt-4o} model achieves the highest labeling accuracy (82.5\%), demonstrating its ability to interpret structured gene expression signatures, and showcasing the feasibility of foundation models for high-throughput annotation in single-cell transcriptomics.}
    \label{fig:marker-label-accuracies}
\end{figure}

\section{Results}

We evaluated DeepSeq's ability to automate structured single-cell annotation using foundation models prompted with top marker genes per cluster. As shown in Figure~\ref{fig:marker-label-accuracies}, our two-stage evaluation assesses both the biological plausibility of marker gene matches and the accuracy of resulting cell-type predictions relative to ground truth annotations.

\subsection{Marker Match Validation}

The top panel in Figure~\ref{fig:marker-label-accuracies} confirms that marker genes extracted for each cluster match canonical gene sets for known cell types, validating the biological grounding of the prompts used for LLM querying. This step ensures that model outputs reflect meaningful transcriptional signatures rather than spurious correlations.

\subsection{LLM Label Accuracy}

We then compared the predicted labels from each LLM against manually curated ground truth. As shown in the bottom panel, the agentic GPT-4o model achieved the highest accuracy (82.5\%), outperforming both earlier GPT-3.5 variants and smaller local models like LLaMA3-1B. These results demonstrate that foundation models, when structured with domain-informed prompts, can approach expert-level annotation performance in high-throughput settings.

\subsection{Reproducibility and Benchmarking}

The full set of results---including per-cluster marker genes, predicted labels, ground truth matches, and evaluation scores---is reproducibly generated via scripts provided in the DeepSeq repository. Each step of the pipeline---from filtering and dimensionality reduction to LLM prompting and evaluation---outputs interpretable logs, enabling precise traceability of every decision made during annotation. This framework supports extensibility to larger datasets, alternative LLM configurations, or modified evaluation strategies.

\begin{figure}[!h]
    \centering
    \includegraphics[width=\linewidth]{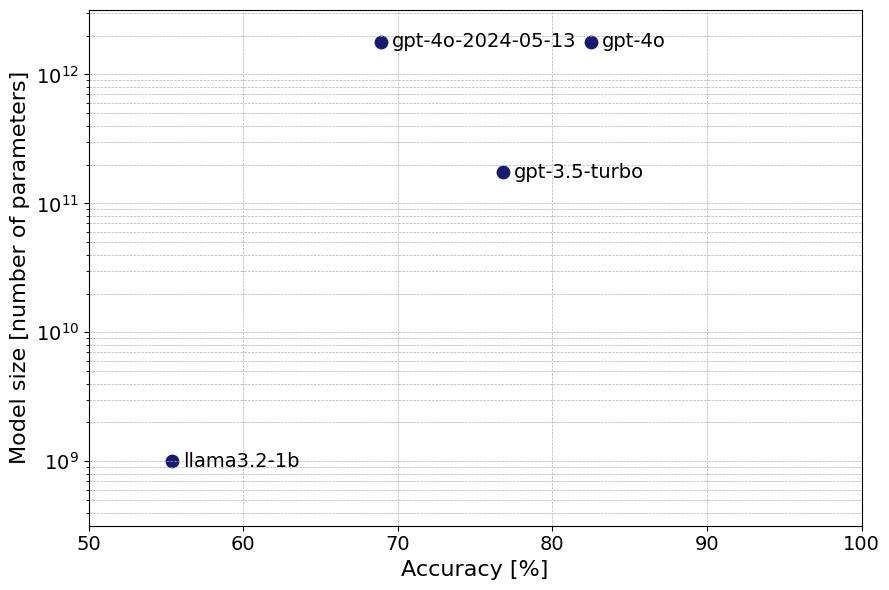}
\caption{Total parameter count versus labeling accuracy for foundation models on structured single-cell data. Accuracy reflects automated cell-type labeling using top marker genes per cluster and prompting strategies derived from \cite{hou2024assessing}. Despite similar active parameter counts, \texttt{gpt-4o} achieves the highest labeling accuracy (82.5\%) while operating within a 1.8T parameter architecture. The logarithmic y-axis reveals how model scale influences annotation performance, highlighting diminishing returns beyond 100B parameters without more domain-specific single-cell data. These trends underscore the need for continued data scaling—toward billions or even trillions of single cells—to approach human-level labeling accuracy.}
    \label{fig:size-accruacy}
\end{figure}

\section{Discussion}

Our results show that foundation models, particularly agentic variants like \texttt{gpt-4o}, can achieve strong performance in structured biological tasks such as cell-type annotation. As illustrated in Figure~\ref{fig:size-accruacy}, GPT-4o achieved 82.5\% agreement with ground truth labels when prompted with top-ranked marker genes per cluster. This level of accuracy, attained without fine-tuning or task-specific supervision, underscores the potential of foundation models for high-throughput biological interpretation.

Interestingly, model performance did not scale linearly with size. The leap in accuracy from LLaMA3-2-1B to GPT-3.5-turbo was larger than the improvement from GPT-3.5-turbo to GPT-4o, despite the latter having significantly more parameters. Since both GPT-3.5 and GPT-4o leverage agentic web search while LLaMA3-2-1B does not, these results suggest that agentic capabilities offer a baseline improvement, but architectural refinements and scaling yield diminishing returns in structured reasoning tasks without more domain-specific data — highlighting the need for experimental generation of such data from high-throughput experiments that will in turn require high-throughput labeling, as demonstrated in this paper. Local models like LLaMA3-1B also performed competitively given their size, reinforcing lightweight deployments in constrained environments.

A key insight from our findings is that, similar to how language models improve with larger parameter counts and more diverse training data, cell-type annotation accuracy also depends on the scale and diversity of experimental input. As shown in Figure~\ref{fig:scaling-projection}, the number of cells profiled in single-cell studies has followed an exponential trajectory, with projections suggesting that datasets containing over $10^9$ cells will become feasible within the decade. This scale is likely necessary to train robust, domain-specific models capable of resolving subtle transcriptional differences across tissues, conditions, and perturbations. We are now at a turning point where the volume of biological data is sufficient to support foundation model–level training and evaluation in structured omics.

Nonetheless, the observed gap between model predictions and perfect label accuracy highlights current limitations in both model capabilities and marker gene distinctiveness. Marker-based prompts are only as informative as the signal contained within each cluster, and foundational models still exhibit brittleness in biologically ambiguous cases. Evaluation scripts in our repository provide insight into these edge cases for reproducible future benchmarking. These findings validate the use of LLMs for structured omics annotation, while motivating design of prompting protocols, marker gene selection, and evaluation pipelines.

\section{Conclusion}

We introduced DeepSeq, a modular pipeline that applies foundation models to the structured domain of single-cell transcriptomics. By using top-ranked marker genes as prompts, DeepSeq enables large language models to perform scalable, automated cell-type labeling with strong agreement to expert-curated ground truth. Our evaluation shows that agentic models equipped with real-time retrieval capabilities outperform static or smaller models, highlighting the importance of model architecture and inference context in structured annotation tasks.

Future work will extend this approach beyond cell-type classification to dynamic biological modeling, including transcriptional perturbation prediction and temporal inference. As single-cell datasets continue to scale, structured prompting combined with model-guided annotation offers a promising foundation for building interpretable, data-driven systems capable of capturing complex biological processes.

These results also suggest that the scaling laws of language models—where performance improves with model size and data—extend to biological annotation. As illustrated in our scaling projections, the exponential growth in single-cell sequencing puts billion-cell datasets within reach. This volume of training data opens the door to training virtual cell foundation models that operate at scale and can generalize across tissue types, organisms, and experimental conditions.

Unlike traditional pipelines constrained by human curation, DeepSeq leverages the compositional reasoning and retrieval capabilities of LLMs to automate annotation with high reproducibility and throughput. As more high-quality single-cell data becomes available, these models will continue to improve. Ultimately, foundation models applied to structured biological data will not only match—but are likely to surpass—human-level annotation performance in both speed and accuracy.

Looking ahead, integrating DeepSeq with multi-omic datasets—such as single-cell ATAC-seq or spatial transcriptomic \cite{wang2025spatialagent}—could further enhance resolving cell identity and state. By extending the prompting framework to handle diverse molecular modalities, DeepSeq can evolve into a general-purpose interface for querying structured biological systems using natural language.

\section*{Acknowledgements}
S.A.A. acknowledges financial support from the Friesecke (1961) Fellowship through the Department of Civil and Environmental Engineering (CEE) at the Massachusetts Institute of Technology (MIT), and is grateful for the advising and mentorship of Professors Heidi Nepf and Ali Jadbabaie. The author thanks members of the AbuGoot Laboratory for insightful discussions, including—but not limited to—Professors Omar Abudayyeh and Jonathan Gootenberg, Thomas Kesheshian, Dr. Juhyung Jung, Elvira Kinzina, Oscar Pitcho, Dan Lesman, Nic Fishman, Jason Lequeyer, Tanush Kumar, and others for their valuable feedback and encouragement. The author also gratefully acknowledges members of the Gladyshev Laboratory, including—but not limited to—Professor Vadim Gladyshev, Dr. Jesse Poganik, Dmitrii Glubokov, and others, for their support.

\section*{Impact Statement}

This paper presents work whose goal is to advance the application of generative AI foundation models for structured biological data, specifically in the context of single-cell transcriptomics. By automating cell-type labeling using agentic prompting strategies, our approach increases annotation throughput and enables scalable deployment in biomedical pipelines. These capabilities have potential implications for diagnostics, perturbation screening, and biological discovery at scale. The methodology developed reflects a broader trend toward integrating generative models with structured biological data, offering a pathway for more versatile and data-driven approaches to life science applications.

As the underlying datasets grow toward billions or trillions of cells, ethical considerations emerge around privacy, model transparency, and equitable generalization. We emphasize the importance of responsible deployment and alignment with expert oversight in real-world health contexts.


\bibliography{paper}
\bibliographystyle{icml2025}

\newpage



\end{document}